# Adversary Model: Adaptive Chosen Ciphertext Attack with Timing Attack

Mohd Anuar Mat Isa[1], Habibah Hashim[2]

Faculty of Electrical Engineering,

40450 UiTM Shah Alam,

Selangor, Malaysia.

[1]anuarls@hotmail.com,
[2]habib350@salam.uitm.edu.my.

*Abstract*—We have introduced a novel adversary model in Chosen-Ciphertext Attack with Timing Attack (CCA2-TA) [1] and it was a practical model because the model incorporates the timing attack. This paper is an extended paper for "A Secure TFTP Protocol with Security Proofs" [1].

*Keywords—Timing Attack, Random Oracle Model, Indistinguishabilit, Chosen Plaintext Attack, CPA, Chosen Ciphertext Attack, IND-CCA1, Adaptive Chosen Ciphertext Attack, IND-CCA2, Trivial File Transfer Protocol, TFTP, Security, Trust, Privacy, Trusted Computing, UBOOT, AES, IOT, Lightweight, Asymmetric, Symmetric, Raspberry Pi.*

## I. INTRODUCTION

This paper is a continuation from our previous work in years 2012 and 2014 [1]–[3]. The paper was written in a general information security terminology with a simple mathematical notation (semi-formal). It was intended for information security practitioner and not for mathematician or cryptographer as the main audience. We hope that this paper will give a worthy understanding of adversary model in Chosen-Ciphertext Attack with Timing Attack and its security proofs. We have considered this paper as a draft paper[1].

## II. RESEARCH GOAL

### A. Objectives

The purpose of this research work is to facilitate a *timing-attack* in the random oracle model. We add the *timing-attack* in an Adaptive Chosen Ciphertext Attacks (CCA2) model.

### B. Motivations

Referring to our previous work [3] [1], we have mentioned the need of a secure TFTP protocol particularly in various network administrative tasks such as monitoring and upgrading of remote embedded device's firmware, where a lightweight protocol such as TFTP is usually employed. The security risks in such situations were also discussed with emphasis on concerns due to physical attacks wherein attackers access and modify Wi-Fi AP hardware and software[4]–[7]. In a preceding work, we proposed an enhanced data communication package for DENX-UBOOT [8] firmware to include a secure TFTP protocol. However, our proposal did not suggest a specific cryptographic protocol for the successful implementation of the secure TFTP protocol [3]. In the effort to further augment the work, a proven secure and practical asymmetric cryptographic scheme, i.e. the Cramer-Shoup (CS) protocol is proposed to be deployed as the underlying cryptographic protocol [9] in the overall scheme. One may refer to [1].

## III. RELATED WORK

### A. Chosen Plaintext Attack (CPA)

Goldwasser-Micali (1982) [10] proposed a *probabilistic encryption* to replace a *trapdoor-function* for a better security evaluation in any encryption scheme. The authors argued that, trapdoor-function do not cover *"the possibility of computing $x$ from $f(x)$ when $x$ is of special form"* and *"the possibility of easily computing some partial information about $x$ from $f(x)$"*[10]. The probabilistic encryption gives a better security property with equal to a probability of flipping a coin such that $\frac{1}{2}$ *head and* $\frac{1}{2}$ *tail* in a fair game. An adversary win in a tossing coin game with probability of $\frac{1}{2} + \varepsilon$; where $\varepsilon$ is an *advantage* that gives the adversary a better chance to win the game.

Goldwasser-Micali proposed an adversary model that gives the adversary a knowledge of encryption protocol and its algorithm, but the adversary cannot obtain any information about a plaintext when given a ciphertext, except that a length of plaintext if an encryption scheme ratio was $x:y$ for $x$ plaintext to $y$ ciphertext. The knowledge of plaintext's length is easy to be obtained by the adversary because of the knowledge of the cryptographic protocol algorithm. To simulate the Goldwasser-Micali's adversary model (CPA), we use indistinguishability test by let an adversary to choose two plaintexts $(m_0, m_1)\ \varepsilon\ \mathcal{M}$ where $(m_0 \neq m_1)$ and $|m_0| = |m_1|$. The plaintexts either $(m_0, m_1)$ is randomly choose to be encrypted using encryption scheme $\Pi = (\mathcal{K}, \mathcal{E}, \mathcal{D})$ by a Challenger:

---

[1] Not complete and not being peer reviewers yet. Everyone is welcome to give any comment/suggestion for further improvements.

$$(pk, sk) \leftarrow \mathcal{K}(1^k)$$
$$b \leftarrow \{0,1\}$$
$$c^* := \mathcal{E}_{pk}(m_b)$$

Let $p(n)$ denote the set of prime in size of $n$, for all sufficiently large $n$.

$$|Pr[success] - Pr[failure]| < \frac{1}{p(n)}$$

A ciphertext $c^*$ is given to the Adversary for a distinguishability test wherein the ciphertext $c^*$ is either encrypted of $(m_0, m_1)$. The encryption scheme $\Pi$ is *semantically secure* if any *probabilistic polynomial-time (PPT)* algorithms that are used in the adversary model to determine a correct plaintext from ciphertext $c^*$ with a negligible probability. Goldwasser-Micali (1984) [11] showed that the *probabilistic encryption* can be implemented under intractability assumption of quadratic residuosity.

### B. Indistinguishability-Chosen Ciphertext Attack (IND-CCA1)

IND-CCA1 is that give an adversary to access decryption function through a decryption oracle. The adversary can ask the oracle to decrypt any ciphertexts $c_i \epsilon \mathcal{C}$ except the one (e.g. $c^*$) that being use for indistinguishability test. This adversary model gives the Adversary more knowledge than CPA's model. However, the decryption oracle can be used by the adversary before the indistinguishability test is happen. Naor-Yung (1990) [12] was the leading that succeed in the IND-CCA1 to prove their public key cryptosystem is secure when the adversary is allow to access decryption oracle before execution of indistinguishability test. The scheme [12] use a non-interactive zero-knowledge (NIZK) with proofs that their protocol satisfied the completeness, soundness and zero-knowledge properties in PPT for a sufficient large $n$ in the $p(n)$.

### C. Indistinguishability-Adaptive Chosen Ciphertext Attack (IND-CCA2)

Rackoff-Simon (1991) [13] argued that an adversary in CCA1 may get an access to a decryption oracle even after the challenge's ciphertext $c^*$ was issued. This attack is a practical security problem because it can be happen in a real-world. A security property for this kind of attack is that prevent the adversary from getting *any useful information from other ciphertext $c_i$* that can helps to get a non-negligible *advantage* to distinguish the challenge's ciphertext $c^*$ in a polynomial time. The authors [13] stressed that it is important to secure against this attack because a digital signature scheme (practical scheme in the real-world) that is vulnerable to this attack. The digital signature scheme is secure "*if any such attacker succeeds in generating a valid signature for this last document with negligible probability*" [13].

IND-CCA2 model is that give an adversary to access decryption function through decryption oracle. The adversary can ask the oracle to decrypt any ciphertext $c_i$ except the one (ciphertext $c^*$) that being use for indistinguishability test.

Referred to Figure 1, the IND-CCA2 allows the Adversary to get a decryption of ciphertext $c_i$ from the oracle in Phase 1 (before) and Phase 2 (after) the challenge messages ($m_0, m_1$ where $|m_0| = |m_1|$ and $m_0 \neq m_1$) are issued to Challenger. The Challenger will choose randomly either $m_0$ or $m_1$ to be encrypted. Ciphertext $c^*$ is send to the Adversary. The Adversary need to distinguish the ciphertext $c^*$ is either $m_0$ or $m_1$ with probability of $\frac{1}{2} + \varepsilon$. If the probability to guess a correct the ciphertext $c^*$ is greater than $\frac{1}{2}$ with non-negligible advantage in PPT; we conclude that the Adversary has an *"advantage"* and the given protocol $\Pi$ is consider not secure in terms of indistinguishability.

$$Advantage_{IND-CCA2}[Adversary, Protocol\ \Pi] =$$
$$|Pr[Experiment(0) = 1] - Pr[Experiment(1) = 1]|\ is\ negligible$$

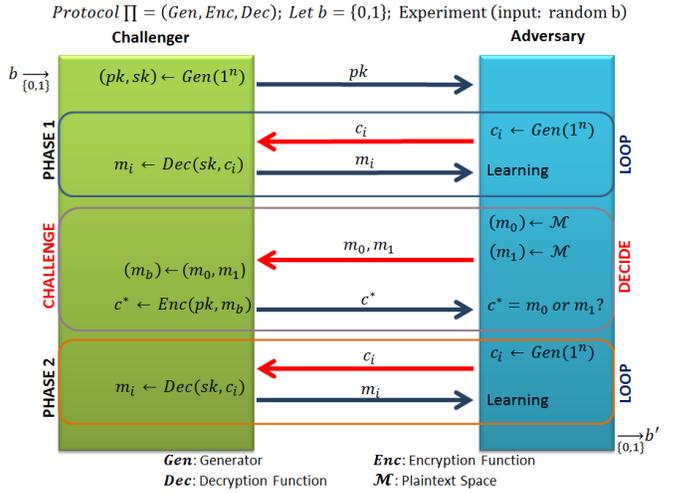

Figure 1: IND-CCA2's Experiment

### IV. PROPOSED SECURITY MODEL

We propose an attack model of Adaptive Chosen-Ciphertext Attack with Timing Attack (CCA2-TA) is that gives an adversary to access identical computing resources in term of computing power (e.g. CPU). The adversary is given the knowledge of time to perform cryptographic computations (e.g. primitive computation and protocol execution) in polynomial time. These were included that the adversary know the delay of network transmission for all transaction in the Phase 1, Phase 2 and Challenge (refer to Figure 2). The adversary is also has the knowledge of CCA2. We defined the proposed security model in general terminology as below:

***Definition 1.0:*** *An Adversary Model CCA2-TA allows an attacker to access runtime for cryptosystem $\Pi = (\mathcal{K}, \mathcal{E}, \mathcal{D})$ to perform $\mathcal{E} \stackrel{\text{def}}{=} encryption$ and $\mathcal{D} \stackrel{\text{def}}{=} decryption$ in a computing machine. The CCA2-TA has the corollary knowledge of CCA2's Adversary Model in the computing machine.*

***Definition 2.0:*** *An Adversary Model CCA2-TA allows an attacker to access runtime for cryptosystem $\Pi = (\mathcal{K}, \mathcal{E}, \mathcal{D})$ to perform $\mathcal{E} \stackrel{\text{def}}{=} encryption$ and $\mathcal{D} \stackrel{\text{def}}{=} decryption$ in a polynomial-time machine using random oracle model. The CCA2-TA has the corollary knowledge of CCA2's Adversary Model for the polynomial-time machine.*

***Theorem 1.0:*** *The cryptosystem $\Pi$ as defined in the Definition 1.0 and 2.0 are secure from the CCA2-TA if the runtime to perform $\mathcal{E}$ and $\mathcal{D}$ are fixed-time $t_{ft}$ for all valid inputs and for all invalid inputs into the function $\mathcal{E}$ and $\mathcal{D}$ with an adversary advantage is negligible. The fixed-time $t_{ft}$ of runtime for function $\mathcal{E}$ to perform encryption and for function $\mathcal{D}$ to perform decryption can be different $t_{ft}(\mathcal{E}) \neq t_{ft}(\mathcal{D})$. The rejected of invalid inputs is returned in fixed-time using that function's runtime.*

***Lemma 1.1:*** *The cryptosystem $\Pi$ as defined in the Definition 2.0 is secure in Indistinguishability CCA2-TA experiment if a fixed-time of function $\mathcal{E}$ is use in issuing a challenge ciphertext $c^*$ in PPT machine with an adversary advantage is negligible. The cryptosystem $\Pi$ can use worse-cases time for all valid inputs in function $\mathcal{E}$ in polynomial-time machine as fixed-time for all execution of function $\mathcal{E}$. This including the rejected of invalid inputs is returned in fixed-time using that function's runtime. The fixed-time using worse-cases time is secure for an implementation [14] but it makes the cryptosystem $\Pi$ running in the most unfit condition.*

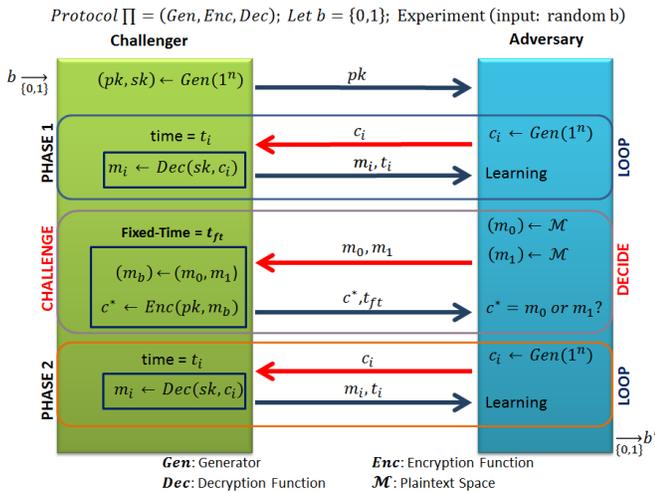

Figure 2: IND-CCA2-TA's Experiment

ACKNOWLEDGMENT

The authors would like to acknowledge the Ministry of Education (MOE) Malaysia for providing the grant 600-RMI/NRGS 5/3 (5/2013), and Universiti Teknologi MARA (UiTM) for supporting this research work.

REFERENCES

[1] Mohd Anuar Mat Isa, Habibah Hashim, Syed Farid Syed Adnan, Jamalul-lail Ab Manan, and Ramlan Mahmod, "A Secure TFTP Protocol with Security Proofs," in Lecture Notes in Engineering and Computer Science: Proceedings of The World Congress on Engineering 2014, (WCE 2014), 2014, vol. 1.

[2] Mohd Anuar Mat Isa, Habibah Hashim, Syed Farid Syed Adnan, Jamalul-lail Ab Manan, and Ramlan Mahmod, "An Experimental study of Cryptography Capability using Chained Key Exchange Scheme for Embedded Devices," in Lecture Notes in Engineering and Computer Science: Proceedings of The World Congress on Engineering 2014, (WCE 2014), 2014.

[3] Mohd Anuar Mat Isa, Nur Nabila Mohamed, Habibah Hashim and R. M. Syed Farid Syed Adnan, Jamalul-lail Ab Manan, "A Lightweight and Secure TFTP Protocol in the Embedded System," in 2012 IEEE Symposium on Computer Applications and Industrial Electronics (ISCAIE 2012), 2012.

[4] Mohd Anuar Mat Isa, Habibah Hashim, Jamalul-lail Ab Manan, Ramlan Mahmod, Mohd Saufy Rohmad, Abdul Hafiz Hamzah, Meor Mohd Azreen Meor Hamzah, Lucyantie Mazalan, Hanunah Othman, Lukman Adnan, "Secure System Architecture for Wide Area Surveillance Using Security, Trust and Privacy (STP) Framework," J. Procedia Eng., vol. 41, no. International Symposium on Robotics and Intelligent Sensors 2012 (IRIS 2012), pp. 480–485, Jan. 2012.

[5] D. D. Clark and D. R. Wilson, "A comparison of commercial and military computer security policies," in IEEE Symposium on Security and Privacy, 1987.

[6] B. Hay, K. Nance, and M. Bishop, "Storm Clouds Rising: Security Challenges for IaaS Cloud Computing," in 2011 44th Hawaii International Conference on System Sciences, 2011, pp. 1–7.

[7] Jamalul-Lail Ab Manan, Mohd Faizal Mubarak, Mohd Anuar Mat Isa, Zubair Ahmad Khattak, "Security, Trust and Privacy – A New Direction for Pervasive Computing," Inf. Secur., pp. 56–60, 2011.

[8] DENX Software Engineering, "DENX U-Boot," 2014. [Online]. Available: http://www.denx.de/wiki/U-Boot/WebHome.

[9] R. Cramer and V. Shoup, "A Practical Public Key Cryptosystem Provably Secure Against Adaptive Chosen Ciphertext Attack," in Lecture Notes in Computer Science: Advances in Cryptology—CRYPTO'98, 1998, pp. 1–18.

[10] S. Goldwasser and S. Micali, "Probabilistic encryption & how to play mental poker keeping secret all partial information," in STOC '82 Proceedings of the fourteenth annual ACM symposium on Theory of computing, 1982, pp. 365–377.

[11] S. Goldwasser and S. Micali, "Probabilistic Encryption," J. Comput. Syst. Sci., vol. 28, no. 2, pp. 270–299, 1984.

[12] M. Naor and M. Yung, "Public-key cryptosystems provably secure against chosen ciphertext attacks," in Proceedings of the twenty-second annual ACM symposium on Theory of computing - STOC '90, 1990, pp. 427–437.

[13] C. Rackoff and D. R. Simon, "Non-Interactive Zero-Knowledge Proof of Knowledge and Chosen Ciphertext Attack," Adv. Cryptol. — CRYPTO '91, vol. LNCS 576, pp. 433–444, 1992.

[14] P. Kocher, "Timing attacks on implementations of Diffie-Hellman, RSA, DSS, and other systems," in Advances in Cryptology—CRYPTO'96, 1996.